\documentclass[11pt]{article}
\def\s{\sigma} 
\def\o{\omega} 
\def\n{\noindent}
\def\r{\ref}
\def\p{\partial}
\def\no{\nonumber}
\begin{document}
\title{{\bf{\Large  Normal ordering and noncommutativity in open bosonic strings}}}
\author{
{\bf {\normalsize Biswajit Chakraborty}\thanks{biswajit@bose.res.in}},\,
 {\bf {\normalsize Sunandan Gan{g}opadhyay}\thanks{sunandan@bose.res.in}} \\
 {\bf {\normalsize Arindam Ghosh Hazra}\thanks{arindamg@bose.res.in}}\\
 {\normalsize S.~N.~Bose National Centre for Basic Sciences,}\\{\normalsize JD Block, Sector III, Salt Lake, Kolkata-700098, India}\\[0.3cm]
}

\maketitle

\begin{abstract}
Noncommutativity in an open bosonic string moving in the presence of a
 background Neveu-Schwarz two-form
field $B_{\mu \nu}$ is investigated in a conformal 
field theory approach, leading
to  noncommutativity at the boundaries. In contrast to several 
discussions, in which boundary conditions are taken as Dirac 
constraints, we first obtain the mode algebra by using the newly
proposed normal ordering, which satisfies both equations of motion and boundary 
conditions. Using these
the commutator among the string coordinates is obtained. Interestingly, 
this new normal ordering  yields the same algebra 
between the modes as the one satisfying only  the equations of motion. 
In this approach, we find that noncommutativity originates more 
transparently and our results match with the existing results in the literature.

\\[0.3cm]

\end{abstract}

\section{Introduction}
Noncommutative(NC) theories has a long history in physics
\cite{sny} and has kindled a lot of interest in the past few years
owing to the inspiration from string theory \cite{sw, dn}. 
Recent progress in string theory indicates that an 
open string moving in the presence
of a background Neveu-Schwarz two form field $B_{\mu \nu}$,
leads to a NC structure \cite{chu, chu1, ard, rb, partha, shir, 
neves, Ors}.
 This structure manifests in 
the noncommutativity in the space-time coordinates of D-branes,
where the end points of the open string are attached. Furthermore, this fact 
reveals the nontrivial role of boundary conditions(BC) in string theory
and the importance of taking them into account when considering
the quantisation of open strings.
Several approaches have been taken to obtain 
such results, for example a Dirac approach \cite{dir} is employed 
with the string BC(s) imposed as  
second class constraints in \cite{ar, br1}.
In a series of recent papers involving bosonic strings \cite{rb},
bosonic membrane  \cite{rbbckk} and superstrings \cite{bcsgaghs},
it has also been shown explicitly
that noncommutativity can be obtained by modifying the canonical 
bracket structure, so that it is compatible with the BC(s). 
This is done in spirit to 
the treatment of Hanson et.al \cite{hrt}, where modified Poisson brackets 
(PBs) were obtained for the free Nambu-Goto string.
Proceeding further Jing and Long \cite{jing}
obtained the PBs among the Fourier components using the
Faddeev-Jackiw symplectic formalism \cite{fj}, so that they are compatible
with these BCs. Using this they obtained the PBs among the open string coordinates revealing the NC structure in the string end points. It is
important to note that all these analysis were essentially confined
to the classical level.
It would thus be quite interesting to study whether this
NC structure can also be obtained from the 
conformal field theory techniques where the analysis is carried out in a 
quantum setting. Indeed, as has been stressed in \cite{sw}
that it is very important to understand this noncommutativity from
different perspectives. More importantly, it is necessary to check explicitly
whether the central charge gets effected by the modified BCs in this case,
as the central charge can be related to the Casimir energy arising from
finite size of the string and is a purely quantum effect \cite{pol}.


It may be recalled in this context that in quantum field theory,
 products of quantum 
fields at the same space-time points are in general singular objects.
 The same thing is true in string theory when one 
multiplies position operators, that can be taken as 
conformal fields on the world sheet.
This situation is well known and one can remove the singular part of the 
operator products by defining normal ordered operators which have
well behaved properties \cite{pol}.
This is important, for example,  when one builds up the generators of conformal 
transformations and investigates the realization of classical symmetries
at quantum level.

Usually normal ordered products of operators are
defined so as to satisfy
the classical equations of motion at quantum level.
However, in a recent paper \cite{br}, new normal ordered 
products have been defined for open string position
operators that additionally satisfy the BCs.
This way one obtains a normal ordering that is also 
valid at string end points. The mathematical problem posed by defining the normal ordering is related  to that 
of calculating Green's functions\cite{AGNY,Schomerus:2002dc,DN,Dolan:2002px}. 
The normal ordered product is defined by subtracting out the corresponding
Green's functions. 
So we can find normal ordered products satisfying 
open string boundary condition 
using the solutions to open string Green's functions.

In this paper, we shall consider the problem of noncommutativity
using this new normal ordering given in \cite{br}.
By using the contour argument and the new $XX$ operator product 
expansion (OPE),we shall find the commutator among the
 Fourier components firstly
and then the commutation relations among string's coordinates
 which reproduces the noncommutative structure obtained in \cite{rb}
and also in \cite{jing}, as mentioned earlier.
We would also like to stress that the above commutator computed
using the $XX$ OPE satisfying the equation of motion
only \cite{pol} (where the modifications due to BCs are not taken into account)
 also leads to the same NC structure.
In contrast to earlier works in this direction as referred
above, this method does not require the BC(s) to be treated as
Dirac constraints, hence the advantage of this method is that one need not
know all the constraint chains by the consistency requirements.
Further the classification of the constraints into the so called primary
or secondary, the first class or the second class is also not required.

The organisation of this paper is as follows. In section 2, we review 
the recent results involving new normal ordered products in \cite{br}
and also give new compact forms of the normal ordered products.
These compact forms differ in structure from those given in \cite{pol},
as the BCs are now taken into account.
Here we also observe that the central charge undergoes no modification,
despite incorporating the BC in normal ordering. This is reassuring, as
this does not affect the critical dimension $D = 26$ of the bosonic string.
In sec 3, we study the symplectic structure of both free and interacting
string using our method (borrowing relevant results from section 2).
Finally, we conclude in section 4.

\section{New Normal ordered products}
Let us first consider the free Polyakov string action,
\begin{equation}
\label{1}
S_{P} \,= -\, 
{ 1 \over 4\pi \alpha^\prime }
\int_{\Sigma} d\tau d\sigma \Big(
G^{ab} \eta_{\mu\nu } \partial_a X^\mu \partial_b X^\nu \,+ \, 
\epsilon^{ab} B_{\mu\nu} \partial_a X^\mu \partial_b X^\nu\Big)
\end{equation}
\noindent 
where $\tau, \sigma$ are the usual world-sheet parameters,
 $G_{ab}$ is the induced world-sheet metric with 
$G_{\tau\tau} = -1, \, G_{\sigma\sigma} = 1$ (upto a Weyl factor) 
in conformal gauge and the antisymmetric
tensor is chosen by $\epsilon^{\tau \s} = 1$. 
$X^\mu (\tau, \sigma)$ are the string coordinates 
in the $D$ dimensional Minkowskian target space with metric
$\eta_{\mu\nu} = (-1, 1, ...., 1)$.

We now make a Wick rotation by defining
$\s^2 = i\tau$   and obtain the classical action
for a bosonic string taking a world-sheet with Euclidean signature:
\begin{equation}
\label{2}
S \,=\, { 1 \over 4\pi \alpha^\prime }
\int_{\Sigma} d^2\sigma \Big(
g^{ab} \eta_{\mu\nu } \partial_a X^\mu \partial_b X^\nu \,+ \,i\, 
\varepsilon^{ab} B_{\mu\nu} \partial_a X^\mu \partial_b X^\nu\Big)
\end{equation}
where  $g^{ab}$ can now be taken to be proportional to the
unit matrix and $\varepsilon^{12} = - \varepsilon^{21} = 1$.
Note that the $D$ dimensional target space-time still has the Lorentzian
signature.

\n
The variation of the action (\r{2}) gives the equation of motion,
\begin{equation}
(\, \partial^2_1 \, +\, \partial^2_2\,)  X^\mu\,=\,0
\label{3}
\end{equation}
and a boundary term that yields the following BCs:
\begin{eqnarray}
\left(\p_1 X^{\mu}(\s^1, \s^2)\,+ \,i\, B_{\mu\nu}
\p_2 X^{\nu}(\s^1, \s^2)\right)\vert_{\s=0, \pi} = 0.
\label{4}
\end{eqnarray}
It is convenient, to introduce complex world sheet coordinates \cite{pol}: 
$ z \,=\,  \s^1 +i \s^2 \, ; \,  \bar{z} \,=\, \s^1 - i \s^2$
and $\p_z = \frac{1}{2}(\p_1 - i\p_2), \, \p_{\bar{z}} = 
\frac{1}{2}(\p_1 + i\p_2)$.

In this notation the action is
\begin{eqnarray}
\label{5}
S \,=\,{ 1 \over 2\pi \alpha^\prime }
\int d^2 z \Big(\p_{z} X^\mu \p_{\bar{z}}X_\mu - B_{\mu \nu}
\p_{z} X^\mu \p_{\bar{z}}X^{\nu}\Big)
\end{eqnarray}
while the classical equations of motion and boundary conditions take the form
\begin{eqnarray}
\partial_{\bar z} \partial_z  X^{\mu}(z, \bar{z}) &=& 0 
\label{6}\\
\left[\eta^{\mu \nu}\Big(\partial_z + \partial_{\bar z}\Big)
- B^{\mu \nu}\Big(\partial_z - \partial_{\bar z}\Big)\right]
X_\nu\vert_{z = -\bar{z},\,  2\pi - \bar{z}} &=& 0
\label{7}
\end{eqnarray}
We now study the properties of quantum operators,
corresponding to the classical variables, by considering the 
expectation values. Defining the expectation value of 
an operator ${\cal F}$ as\cite{pol}:
\begin{eqnarray}
\langle {\cal F} [X] \rangle \,=\,\int [dX] \exp ( - S[X] ) {\cal F} [X] 
\label{28}
\end{eqnarray}
\noindent 
and using the fact that the path integral of a total derivative vanishes one
finds:
\begin{eqnarray}
0 &=& \int [dX] {\delta \over \delta X^\nu (z^\prime , {\bar z}^\prime )}
exp ( - S[X] ) \,=\,
\Big\langle \,{1\over \pi\alpha^\prime } \partial_{\bar z^\prime}
\partial_{z^\prime} X_\nu ( z^\prime, \bar z^\prime ) \Big\rangle \nonumber\\
&+& {1\over 2\pi \alpha^\prime}  \oint_{_{\partial \Sigma}} \delta^2 (z - z^\prime) 
\Big\langle \Big( \eta_{\nu\mu} ( \partial_z + \partial_{\bar z} ) + 
 B_{\nu\mu} ( \partial_z - \partial_{\bar z} ) \Big) X^\mu (z, {\bar z} ) 
 \,\Big\rangle dz.
\label{29}
\end{eqnarray}
\noindent 
The last (singular) term is integrated over the boundary, 
where $dz = - d \bar z$. We thus find that this equation
 implies that both string equations of motion 
and boundary condition hold as expectation values.
So the corresponding quantum position operators ${\hat X}^\mu$ 
(in target space) satisfy (as long as they are not 
multiplied by other local operators coincident at 
the same world-sheet point) the following equations:
\begin{eqnarray}
\label{30a}
\partial_{\bar z} \partial_{z} {\hat X}^\nu ( z, \bar z ) &=& \,0  \\
\Big( \eta_{\nu\mu} ( \partial_z + \partial_{\bar z} ) - 
 B_{\nu\mu} ( \partial_z - \partial_{\bar z} ) \Big) {\hat X}^{\mu}
\vert_{_{z = -\bar z,\, 2\pi - \bar z }} &=& 0 
\label{30}
\end{eqnarray}
which are nothing but the quantum version of (\r{6}, \r{7}).
Proceeding in the same way, we can consider a pair of local 
operators which may now be coincident to show that their products 
 at the quantum level satisfy\cite{br} :
\begin{equation}
\partial_{\bar z^\prime} \partial_{z^\prime} {\hat X}^\mu ( z^\prime, {\bar z}^\prime) 
{\hat X} ^\nu (z^{\prime\prime}, {\bar z^{\prime\prime}} ) = \,- \pi \alpha^\prime\,
\eta^{\mu\nu} 
\, \delta^2
( z^\prime - z^{\prime\prime} , {\bar z}^\prime - {\bar z}^{\prime\prime} )
\end{equation}
\begin{equation}
\Big( \eta_{\nu\mu} ( \partial_{z^\prime} + \partial_{\bar z^\prime} ) - 
 B_{\nu\mu} ( \partial_{z^\prime} - \partial_{\bar z^\prime} ) \Big) 
{\hat X}^\mu ( z^\prime, {\bar z}^\prime)
{\hat X}^\rho ( z^{\prime\prime}, {\bar z^{\prime\prime}} )
\vert_{_{\mathrm{Bound}.}} \,=\, 0
\end{equation}
where we have considered a pair of local operators.
Now if we introduce the operation of normal ordering in the 
standard way\cite{pol},
\begin{eqnarray}
:\,{\hat X}^\mu (z, \bar z )\, :\, &=& \, {\hat X}^\mu (z, \bar z )\no\\
:\,{\hat X}^\mu (z, \bar z )\, {\hat X}^\nu (z^\prime, {\bar z}^\prime)\,:
\, &=& \, {\hat X}^\mu (z , \bar z  ) \,{\hat X}^\nu (z^\prime ,{\bar z}^\prime)\, + 
{ \alpha^\prime \over 2} \eta^{\mu\nu} {\mathrm{ln}} \vert z - z^\prime \vert^2
\label{9}
\end{eqnarray} 
it satisfies the equation of motion (\r{6}) at the quantum level, 
but fails to satisfy the boundary conditions (\r{7}). In \cite{br},
the authors have introduced a different kind of normal 
ordered product satisfying both equation of motion and boundary conditions.

At this point it is more convenient to choose world sheet 
coordinates, related to these $z$ coordinates by conformal
transformation, that simplify the representation of the boundary,
\begin{eqnarray}
\o\, = \,\mathrm{exp}\left(-iz\right)\, =\, e^{- i \s^1 + \s^2} \,\,\,;\,\,\,{\bar \o} = e^{i \s^1 + \s^2}. 
\label{10}
\end{eqnarray} 
In this present coordinates the complete boundary corresponds 
just to the region $\o = {\bar \o}$.
On the other hand, the action (\r{5})  along with equation of motion (\r{30a})
in terms of $\o, {\bar \o}$ has still the same form, while 
the form of BCs are slightly altered:
\begin{eqnarray}
\label{31}
\p_{\bar{\o}}\p_{\o} {\hat X}^\mu(\o, \bar{\o}) &=& 0 \\
\Big( \eta_{\mu\nu} ( \partial_\o - \partial_{\bar \o} ) - 
 B_{\mu\nu} ( \partial_\o + \partial_{\bar \o} ) \Big) 
{\hat X}^\nu\vert_{_{\o = \bar \o}} \,&=&\, 0\,\,.
\label{31q}
\end{eqnarray}
The corresponding new normal ordering is given by \cite{br}:
\begin{eqnarray}
\label{11}
\mbox{{\bf :}} \, {\hat X}^{\mu}(\o, \bar \o )\, {\hat X}^{\nu}(\o^\prime, {\bar \o}^\prime)
\,\mbox{{\bf :}} &=&
 {\hat X}^{\mu}(\o, \bar \o ) \, {\hat X}^{\nu}(\o^\prime, {\bar \o}^\prime) 
+ \frac{\alpha'}{2}
\eta^{\mu \nu} \mathrm{ln} \vert \o-\o^\prime \vert^2  
+ \frac{\alpha'}{2}
\left( [ \eta - B ]^{-1}\,[ \eta + B ] \right)^{\mu \nu}\no \\ 
&& \mathrm{ln}\left(\o - \bar{\o}^{\prime}\right) +  \frac{\alpha'}{2} 
\Big( [ \eta  - B ] \, [ \eta + B ]^{-1} \,\Big)^{\mu \nu} \,\, 
\mathrm{ln} (\bar{\o}- \o^{\prime} )
\end{eqnarray}
which satisfy both equation of motion and open string 
BCs (\r{31}, \r{31q}) at the quantum level. These additional terms 
can be understood easily as `image' contribution as in electrostatics.

Now for any arbitary functional ${\cal{F}}[X]$, the new 
 normal ordering (in absence of the $B$ field)
can be compactly written as:
\begin{eqnarray}
\label{32}
\mbox{{\bf :}}\,{\cal{F}}\,\mbox{{\bf :}} = \exp\left(
{\alpha^\prime \over 4} \int d^2 \o_1 d^2 \o_2\, \left[\mathrm{ln} 
\vert \o_1 - \o_2\vert^2 + \mathrm{ln} \vert 
\o_1 - \bar{\o}_2 \vert^2\right]\frac{\delta}{\delta X^{\mu}
(\o_1, \bar{\o}_1)}\,\frac{\delta}{\delta X_{\mu}
(\o_2, \bar{\o}_2)}\right){\cal{F}}.
\end{eqnarray}
For example, this reproduces correctly the expression given in (\r{11}),
as one can easily verify for $B=0$.\\
The OPE for any pair of operators, satisfying the BCs, can be generated from
\begin{eqnarray}
\label{33}
\mbox{{\bf :}}\,{\cal{F}}\,\mbox{{\bf :}} \, 
\mbox{{\bf :}}\,{\cal{G}}\,\mbox{{\bf :}}= \exp\left(
{\alpha^\prime \over 4} \int d^2 \o_1 d^2 \o_2\, \left[\mathrm{ln} 
\vert \o_1 - \o_2\vert^2 + \mathrm{ln} \vert 
\o_1 - \bar{\o}_2 \vert^2\right]\frac{\delta}{\delta X^{\mu}
(\o_1, \bar{\o}_1)}\,\frac{\delta}{\delta X_{\mu}
(\o_2, \bar{\o}_2)}\right)\mbox{{\bf :}}\,{\cal{F}}\, {\cal{G}}\,\mbox{{\bf :}}.
\end{eqnarray}
It is now easy to verify that the $TT$ OPE involving
energy-momentum tensor
\begin{eqnarray}
\label{32q}
T(\o) = - \frac{1}{\alpha^\prime} \mbox{{\bf :}}\,
\p X^{\mu}(\o) \p X_{\mu}(\o) \,\mbox{{\bf :}}
\end{eqnarray}
undergoes no modification.
Indeed, using the above definition we obtain the following OPE:
\begin{eqnarray}
\label{41}
\mbox{{\bf :}}\,\p X^\mu(\o)\, \p X_\mu(\o)\,\mbox{{\bf :}}\, 
\mbox{{\bf :}}\,\p^{\prime}X^\nu(\o^{\prime})\,  
\p^{\prime}X_\nu(\o^{\prime})\,\mbox{{\bf :}}
&=& \mbox{{\bf :}}\,\p X^\mu(\o)\, \p X_\mu(\o)\,\p^{\prime}X^\nu(\o^{\prime})\,
\p^{\prime}X^\nu(\o^{\prime})\,\mbox{{\bf :}} \no \\
&& -4 \cdot\frac{\alpha^\prime}{2}\left(\p\p^{\prime}\mathrm{ln}\vert
\o - \o^{\prime}\vert^2\right)\mbox{{\bf :}}\,\p X^\mu(\o)\,\p^{\prime}X_\nu(\o^{\prime})\,\mbox{{\bf :}} \no \\
&& +2 \cdot \delta^{\mu}_{\ \mu}\left(\frac{\alpha^{\prime}}{2}
\p\p^{\prime}\mathrm{ln}\vert \o - \o^{\prime}\vert^2\right)^2 \no \\
&\sim & \frac{D\alpha^{\prime 2}}{2\left(\o - \o^{\prime}\right)^4}
- \frac{2\alpha^{\prime}}{\left(\o - \o^{\prime}\right)^2}\mbox{{\bf :}}\,
\p^{\prime}X^\mu(\o^{\prime})\, \p^{\prime}X_\mu(\o^{\prime})
\,\mbox{{\bf :}}\no \\
&& - \frac{2\alpha^{\prime}}{\left(\o - \o^{\prime}\right)}
\p^{\prime 2}X^\mu(\o^{\prime})\, \p^{\prime}X_\mu(\o^{\prime})
\,\mbox{{\bf :}}
\end{eqnarray}
where $\sim$ mean `equal upto nonsingular terms'. The above result
is same as that of (\cite{pol}) which is obtained by using the usual
normal ordering satisfying the equation of motion only.
This also implies that the Virasoro algebra remains the
same as that of (\cite{pol}). So the new normal ordering (\r{11})
(with $B= 0$) has no impact on the central charge.

We shall make use of the results discussed here in the next section
where we study both free and interacting open bosonic strings.
\section{Mode expansions and Non-Commutativity for bosonic strings} 
\subsection{Free open strings} 
In this section, we consider the mode expansions of free 
($B_{\mu \nu} = 0$) bosonic strings.
We start with the closed string first.
In the $X^{\mu}$ theory (\r{5}), $\p X^{\mu}$ and $\bar{\p} X^{\mu}$ are
(anti)holomorphic and so have the following Laurent expansions,
\begin{eqnarray}
\label{12}
\p X^{\mu}(\o) &=& -i \left(\frac{\alpha^{\prime}}{2}\right)^{\frac{1}{2}}
\sum^{\infty}_{m= -\infty} \frac{\alpha^{\mu}_{m}}{\o^{m+1}}\no \\
 \bar{\p} X^{\mu}(\bar{\o}) &=& -i \left(\frac{\alpha^{\prime}}{2}
\right)^{\frac{1}{2}}
\sum^{\infty}_{m= -\infty} \frac{{\tilde{\alpha}}^{\mu}_{m}}{\bar{\o}^{m+1}}.
\end{eqnarray}
Now the Neumann BC(s) (\r{31q}) in case of free open strings 
(i.e. $B_{\mu\nu} = 0$) requires
$\alpha = \tilde{\alpha}$ in the expansions 
(\r{12}). The expansion for $X^{\mu}$ is then:
\begin{eqnarray}
\label{13}
X^{\mu}(\o, \bar{\o}) = x^{\mu} -i\alpha^{\prime}p^{\mu}
\mathrm{ln}\vert \o \vert^2 + i \left(\frac{\alpha^{\prime}}{2}
\right)^{\frac{1}{2}}\sum_{m \neq 0}\frac{\alpha^{\mu}_{m}}{m}
\left( \o^{-m} + \bar{\o}^{-m}\right)
\end{eqnarray}
where, $x^{\mu}$ and $p^{\mu} = \frac{1}{\sqrt{2 \alpha^{\prime}}}
\alpha^{\mu}_{0}$ are the centre of mass coordinate and 
momentum respectively.

Now the expressions (\r{12}) for open strings can be equivalently written as:
\begin{eqnarray}
\alpha^{\mu}_{m} &=& \left(\frac{2}{\alpha^{\prime}}\right)^{\frac{1}{2}}
\oint \frac{d\o}{2\pi}\, \o^m \p X^{\mu}(\o)\, = \,\oint \frac{d\o}{2\pi i}\, j^{\mu}_{m}(\o)
\label{38}
\end{eqnarray}
where, $j^{\mu}_{m}(\o) = \sqrt{\frac{2}{\alpha^{\prime}}}
 i\, \o^m \p X^{\mu}(\o)$ is the corresponding holomorphic current. 
The commutation relation between $\alpha$'s can be worked out
from the contour argument and the $X\, X$ OPE\footnote{
Note that  here we have used the new normal ordering 
(\r{11}) for free open string, yet
the commutation relations (\r{14}) remain same 
(see \cite{pol}).}\cite{pol},
\begin{eqnarray}
\label{14}
\left[\alpha^{\mu}_{m}, \alpha^{\nu}_{n}\right] &=& 
\oint \frac{d\o_2}{2\pi i}\, \mathrm{Res}_{\o_1 \to \o_2}\left(
j^{\mu}_{m}(\o_1)\,j^{\nu}_{n}(\o_2)\right) \no \\
&=& m \, \delta_{m, -n}\eta^{\mu \nu} 
\end{eqnarray}
At this stage, it should be noted that the above approach
 does not give the algebra
among the zero modes, i.e. the centre of mass variables 
$[x^\mu, p^\nu]$ and $[x^\mu, x^\nu]$ of the open string.
However, the results can be
derived using standard techniques (as has been done in 
\cite{jing}) and read,
\begin{eqnarray}
\left[x^{\mu}, p^{\nu}\right] &=& i \eta^{\mu \nu}.
\label{14a}
\end{eqnarray}
The conjugate momenta $\Pi_{\mu} = \frac{1}{2\pi\, 
\alpha^{\prime}} \dot{X^{\mu}}$ 
corresponding to $X^{\mu}$
can be calculated from the action (\r{1}) which has a Lorentzian
signature for the world-sheet. In order to make a transition to Euclidean
world-sheet, we make use of Wick rotation as before, 
by defining $\s^2 = i \tau$, so that $\dot{X^{\mu}} = i 
\frac{\p X^{\mu}}{\p \s^2}$. The $\Pi^{\mu}(\s^1, \s^2)$ can be recast as
a function of $\o$ and $\bar{\o}$ using (\r{10}), so that its mode
expansion becomes:
\begin{eqnarray}
\Pi^{\mu}(\o, \bar{\o}) = \frac{1}{2\pi\, \alpha^{\prime}}
\left[2 \alpha^{\prime} p^{\mu} + \left(\frac{\alpha^{\prime}}{2}
\right)^{\frac{1}{2}} \sum_{m \neq 0}\alpha^{\mu}_{m} 
\left( \o^{-m} + \bar{\o}^{-m}\right)\right]
\label{15}
\end{eqnarray}
where we have made use of the mode expansion of 
$X^{\mu}(\o, \bar{\o})$ (\r{13}).
The commutation relations between $X^{\mu}(\o, \bar{\o})$ and 
$\Pi^{\nu}(\o^{\prime}, \bar{\o^{\prime}})$ are then obtained by using
(\r{14}, \r{14a}) as,
\begin{eqnarray}
\left[X^{\mu}(\o, \bar{\o}), \Pi^{\nu}(\o^{\prime}, 
\bar{\o^{\prime}})\right] = i \eta^{\mu \nu}\left(\frac{1}{\pi}
+ \frac{1}{4\pi} \sum_{m \neq 0}\left(\o^{-m} + \bar{\o}^{-m}\right)
\left( \o^{\prime m} + \bar{\o^{\prime}}^{m}\right)\right).
\label{16}
\end{eqnarray}
To obtain the usual equal time (i.e. $\tau = \tau^{\prime}$)
commutation relation we first rewrite (\r{16}) in ``$z$ frame''
using (\r{10}) and then in terms of $\s^1, \ \s^2$ to find,
\begin{eqnarray}
\left[X^{\mu}(\s^1, \s^2), \Pi^{\nu}(\s^{1\,\prime}, \s^{2\,\prime})\right]
= i \eta^{\mu \nu}\left[\frac{1}{\pi}
+ \frac{1}{\pi} \sum_{m \neq 0} \exp^{-m\left(\s^2 -
\s^{2\,\prime}\right)} \mathrm{cos}\left(m \s^1\right) 
\mathrm{cos}\left(m \s^{\prime\,1}\right)\right].
\label{17}
\end{eqnarray}
Finally substituting $\tau = \tau^{\prime}$ i.e. $\s^2 = \s^{2\,\prime}$
and $\s^1 = \s$ we get
back the usual equal time commutation relation,
\begin{eqnarray}
\left[X^{\mu}(\tau, \s), \Pi^{\nu}(\tau, \s^{\prime})\right]
= i \eta^{\mu \nu} \Delta_{+}\left(\s , \s^{\prime}\right)
\label{18}
\end{eqnarray}
where, 
\begin{eqnarray}
\Delta_{+}\left(\s , \s^{\prime}\right) = \left[\frac{1}{\pi}
+ \frac{1}{\pi} \sum_{m \neq 0} \mathrm{cos}\left(m \s\right) 
\mathrm{cos}\left(m \s^{\prime}\right)\right].
\label{18aa}
\end{eqnarray}
It is easy to see that (\r{18}) is compatible with Neumann BCs and 
reproduces the result in \cite{rb, bcsgaghs, hrt, jing}.
\subsection{Open string in the constant B-field background} 
We now analyse the open string moving in presence of a background
Neveu-Schwarz two form field $B_{\mu \nu}$. 
To begin with, let us again consider the Laurent expansion of $\p X^{\mu}(\o)$
and $\bar{\p} X^{\mu}(\bar{\o})$ (\r{12}) for the case of closed
string. But now the corresponding open string Laurent expansion is 
obtained by imposing the BCs, given in (\r{31q}) with $B_{\mu \nu} \neq 0$
consequently, the modes $\alpha$ and $\tilde{\alpha}$ now satisfy:
\begin{eqnarray}
\label{34}
\alpha^{\mu}_{m} - B^{\mu}_{\ \nu}\alpha^{\nu}_{m} = 
\tilde{\alpha}^{\mu}_{m} + B^{\mu}_{\ \nu}\tilde{\alpha}^{\nu}_{m}.
\end{eqnarray}
So there exists only one set of independent modes 
$\gamma^{\mu}_{m}$, which can be thought of as the modes of free
strings and is related to $\alpha^{\mu}_{m}$ and $\tilde{\alpha}^{\mu}_{m}$ by: 
\begin{eqnarray}
\label{35}
\alpha^{\mu}_{m} &=& \left(\delta^{\mu}_{\ \nu} + B^{\mu}_{\ \nu}
\right)\gamma^{\nu}_{m} := \left[({1\!\mbox{l}} + B)\gamma
\right]^{\mu}_{m} \no \\
\tilde{\alpha}^{\mu}_{m} &=& \left(\delta^{\mu}_{\ \nu} - B^{\mu}_{\ \nu}
\right)\gamma^{\nu}_{m} := \left[({1\!\mbox{l}} - B)\gamma\right]^{\mu}_{m} .
\end{eqnarray}
Note that under world-sheet parity transformation 
$\alpha^{\mu}_{m} \leftrightarrow \tilde{\alpha}^{\mu}_{m}$, as $B_{\mu \nu}$
is a world-sheet pseudo-scalar.  
Substituting (\r{35}) in (\r{12}), we obtain  the 
following Laurent expansions for $\p X^{\mu}$ and $\bar{\p} X^{\mu}$:
\begin{eqnarray}
\label{19}
\p X^{\mu}(\o) &=& -i \left(\frac{\alpha^{\prime}}{2}\right)^{\frac{1}{2}}
\sum^{\infty}_{m= -\infty} \frac{\left[({1\!\mbox{l}} + B)
\gamma\right]^{\mu}_{m}}{\o^{m+1}} \\
\bar{\p} X^{\mu}(\bar{\o}) &=& -i \left(\frac{\alpha^{\prime}}{2}
\right)^{\frac{1}{2}}
\sum^{\infty}_{m= -\infty} \frac{\left[({1\!\mbox{l}} - B)
\gamma\right]^{\mu}_{m}}{\bar{\o}^{m+1}}\no.
\end{eqnarray}
Integrating the expansion (\r{19}) we obtain the mode expansion
of $X^{\mu}(\o, \bar{\o})$ for the interacting string:
\begin{eqnarray}
\label{20}
X^{\mu}(\o, \bar{\o}) &=& x^{\mu} -i\alpha^{\prime}p^{\mu}
\mathrm{ln}\vert \o \vert^2 - i \alpha^{\prime} B^{\mu}_{\ \nu}p^{\nu}
\left(\mathrm{ln}\o - \mathrm{ln}\bar{\o}\right) \no \\
&&+ \, i \left(\frac{\alpha^{\prime}}{2}
\right)^{\frac{1}{2}}\sum_{m \neq 0}\left[\frac{\gamma^{\mu}_{m}}{m}
\left( \o^{-m} + \bar{\o}^{-m}\right) + \frac{1}{m}B^{\mu}_{\ \nu}\,
\gamma^{\nu}_{\ m}\left( \o^{-m} - \bar{\o}^{-m}\right)\right].
\end{eqnarray}
Now the expressions (\r{19}) for open interacting strings can 
also be equivalently written as:
\begin{eqnarray}
\left[({1\!\mbox{l}} + B)\gamma\right]^{\mu}_{m} &=& 
\left(\frac{2}{\alpha^{\prime}}\right)^{\frac{1}{2}}
\oint \frac{d\o}{2\pi}\, \o^m \p X^{\mu}(\o) \no \\
\left[({1\!\mbox{l}} - B)\gamma\right]^{\mu}_{m} &=& 
- \left(\frac{2}{\alpha^{\prime}}\right)^{\frac{1}{2}}
\oint \frac{d\bar{\o}}{2\pi}\, \bar{\o}^m \p X^{\mu}(\bar{\o}).
\label{39}
\end{eqnarray}
The commutation relation between $\gamma$'s can be obtained
from the contour argument (using (\r{39})) and the $X\, X$ OPE (\r{11}):
\begin{eqnarray}
\label{22}
\left[\gamma^{\mu}_{m}, \gamma^{\nu}_{n}\right] 
= m\, \delta_{m, -n}\left[\left({1\!\mbox{l}}- B^2\right)^{-1}\right]^{\mu \nu}
=  m\, \delta_{m, -n} \left(M^{-1}\right)^{\mu \nu}
\end{eqnarray}
where, $M = ({1\!\mbox{l}} - B^2)$ ; $(B^2)^{\mu \nu} = B^{\mu}_{\ \rho}B^{\rho \nu}$\footnote{Here we should note that $\left({1\!\mbox{l}}\right)^{\mu \nu} 
= \eta^{\mu \nu}$.}. 
Once again the algebra
among the zero modes i.e. the centre of mass variables 
$[x^\mu, p^\nu]$ and $[x^\mu, x^\nu]$ of the open string cannot be obtained from the above contour arguements.
The results can be derived using standard techniques (as 
discussed earlier) and read \cite{jing},
\begin{eqnarray}
\label{23}
\left[x^{\mu}, p^{\nu}\right] &=& i \left(M^{-1}\right)^{\mu \nu} \no \\
\left[x^{\mu}, x^{\nu}\right] &=& -2i\,\alpha^\prime
\pi \left(M^{-1}B\right)^{\mu \nu}.
\end{eqnarray}
Now proceeding as in the free case, the conjugate momentum 
$\Pi^{\mu}(\o. \bar{\o})$ corresponding to (\r{20}) is:
\begin{eqnarray}
\label{27}
\Pi^{\mu}(\o, \bar{\o}) &=& \frac{1}{2\pi \alpha^{\prime}}
\left[2 \alpha^{\prime}\, M^{\mu \rho}\,p_{\rho}
+ \left(\frac{\alpha^{\prime}}{2}\right)^{\frac{1}{2}} 
\sum_{m \neq 0}M^{\mu}_{\ \rho}\,\gamma^{\rho}_{\ m}
\left( \o^{-m} - \bar{\o}^{-m}\right)\right].
\end{eqnarray}
The commutators among the canonical variables $X^{\mu}(\o, \bar{\o}), 
\Pi^{\nu}(\o, \bar{\o})$ can be computed by
using (\r{22}), (\r{23}),
\begin{eqnarray}
\label{24}
\left[X^{\mu}(\o, \bar{\o}), \Pi^{\nu}(\o^{\prime}, 
\bar{\o^{\prime}})\right] &=& i \eta^{\mu \nu}\left(\frac{1}{\pi}
+ \frac{1}{4\pi} \sum_{m \neq 0}\left(\o^{-m} + \bar{\o}^{-m}\right)
\left( \o^{\prime m} + \bar{\o^{\prime}}^{m}\right)\right. \no \\
&& \left. + \frac{1}{4\pi}\sum_{m \neq 0}B^{\mu \nu}
\left(\o^{-m} - \bar{\o}^{-m}\right)
\left( \o^{\prime m} + \bar{\o^{\prime}}^{m}\right)\right)
 \\
\left[X^{\mu}(\o, \bar{\o}), 
X^{\nu}(\o^{\prime}, \bar{\o^{\prime}})\right] &=& \alpha^{\prime}
\left(M^{-1}\right)^{\mu \nu}\left(\mathrm{ln}\vert \o^{\prime}\vert^2
- \mathrm{ln}\vert \o \vert^2\right) - \alpha^{\prime}\left(M^{-1} B
\right)^{\mu \nu}\left(\mathrm{ln}\frac{\o^{\prime}}{\bar{\o^{\prime}}}
+ \mathrm{ln}\frac{\o}{\bar{\o}} + 2 i\, \pi\right) \no \\
&& + \frac{\alpha^{\prime}}{2}\left(\sum_{m \neq 0}\frac{1}{m}\left[
\left(M^{-1}\right)^{\mu \nu}\left(\o^{-m} + \bar{\o}^{-m}\right) 
\left( \o^{\prime m} + \bar{\o^{\prime}}^{m}\right)\right.\right. \no \\
&&\left.\left.+ B^{\mu}_{\ \rho}\, B^{\nu}_{\ \s}\left(
M^{-1}\right)^{\mu \nu}\left(\o^{-m} - \bar{\o}^{-m}\right) 
\left( \o^{\prime m} - \bar{\o^{\prime}}^{m}\right)\right.\right. \no \\
&&\left.\left. - \left(
M^{-1}B\right)^{\mu \nu}\left(\o^{-m} + \bar{\o}^{-m}\right) 
\left( \o^{\prime m} - \bar{\o^{\prime}}^{m}\right)
\right.\right. \no \\
&&\left.\left. + \left(
M^{-1}B\right)^{\mu \nu}\left(\o^{-m} + \bar{\o}^{-m}\right) 
\left( \o^{\prime m} + \bar{\o^{\prime}}^{m}\right)
\right]\right)\no \\
\left[\Pi^{\mu}(\o, \bar{\o}), \Pi^{\nu}(\o^{\prime}, 
\bar{\o^{\prime}})\right] &=& 0 \no.
\end{eqnarray}
Now proceeding as before, we can rewrite the above 
commutation relation in $\s^1, \s^2$ coordinates to obtain the following,
\begin{eqnarray}
\label{25}
\left[X^{\mu}(\s^1, \s^2), \Pi^{\nu}(\s^{1\prime}, \s^{2\prime})\right] 
&=& i \eta^{\mu \nu}\left(
\frac{1}{\pi}
+ \frac{1}{\pi} \sum_{m \neq 0} \exp^{-m\left(\s^2 -
\s^{2\,\prime}\right)}\left[ \mathrm{cos}\left(m \s^1\right) 
\mathrm{cos}\left(m \s^{\prime\,1}\right)
\right.\right. \no \\
&& \left.\left. + B^{\mu \nu}
\mathrm{sin}\left(m \s^1\right) 
\mathrm{cos}\left(m \s^{\prime\,1}\right)\right]\right) \\
\left[X^{\mu}(\s^1, \s^2), 
X^{\nu}(\s^{1\prime}, \s^{2\prime})\right] &=& 2\alpha^{\prime}
\left(M^{-1}\right)^{\mu \nu}\left(\s^{2\prime} - \s^{2}\right)
+2i \alpha^{\prime}\left(M^{-1} B
\right)^{\mu \nu}\left(\s^{1 \prime} + \s^{1} - \pi\right) \no \\
&& + 2\alpha^{\prime}\left(\sum_{m \neq 0}\frac{1}{m}e^{-m\left(\s^{2} 
- \s^{2\prime}\right)}\left[
\left(M^{-1}\right)^{\mu \nu} 
\mathrm{cos}(m\s^1)\mathrm{cos}(m\s^{1\prime})\right.\right. \no \\
&&\left.\left.+ B^{\mu}_{\ \rho}\, B^{\nu}_{\ \s}\,
\mathrm{sin}(m\s^1)\, \mathrm{sin}(m\s^{1\prime})
\right.\right. \no \\
&&\left.\left. +i \left(M^{-1}B\right)^{\mu \nu}
\mathrm{sin} \left(m(\s^1 + \s^{1\prime})\right)\right]\right)\no .
\end{eqnarray}
Finally we obtain the equal time commutation relations by
identifying $\tau = \tau^{\prime}$, i.e. $\s^2 = \s^{2\,\prime}$
and setting $\s^1 = \s$,
\begin{eqnarray}
\label{26}
\left[X^{\mu}(\tau, \s), 
\Pi^{\nu}(\tau, \s^{\prime})\right] &=& i\eta^{\mu \nu}
\Delta_{+}(\s, \s^{\prime})\no \\
\left[X^{\mu}(\tau, \s), 
X^{\nu}(\tau, \s^{\prime})\right] &=& 2i \alpha^{\prime}
\left(M^{-1}B\right)^{\mu \nu}\left[\s + \s^{\prime} - \pi + \sum_{n \neq 0}
\frac{1}{n}\mathrm{sin}\, \left(n(\s + \s^{\prime})\right)\right].
\end{eqnarray}
One can explicitly check that these commutators are
compatible with BCs and reproduces the result in (\cite{chu, chu1, rb, br1,
jing}).
\section{Conclusion}
In this paper, we have used conformal field theory techniques
to compute the commutator among Fourier components which is unlike 
the method in (\cite{jing}), where the algebra among the Fourier components
have been computed using the Faddeev-Jackiw symplectic formalism. This is then 
used to obtain the commutator between the basic fields. 
The advantage of this approach is that the results one obtains takes
into account the quantum effects right from the beginning,
in contrary to the previous 
investigations, which were made essentially at the classical level 
\cite{chu, chu1, rb, br1, jing},
so that the question of the existence of quantum effects, if any,
can be addressed immediately. For example, it has been checked 
that the new normal ordering, as proposed in \cite{br}, which takes into
account the BCs has no bearing on the central charge in case of free
bosonic string. 
Finally, we also computed the oscillator algebra in presence of the $B$
field which is a parity-odd field on the string world-sheet. Consequently
in presence of this $B$ field, the left and right moving modes appearing in the  Laurent series expansions of the  (anti)holomorphic fields $\p X^{\mu}$ and $\bar{\p} X^\mu$ (\r{12}) of the closed string are 
no longer equal when open string BCs (\r{31q}) are imposed to obtain the
corresponding Laurent expansions. These rather get
 related to the free oscillator
modes $\gamma^{\mu}_{m}$ (\r{35}) in a parity asymmetric way.
Using these expressions of left and right moving modes, we rewrite
the (anti)holomorphic fields $\p X^{\mu}$ and $\bar{\p} X^\mu$ entirely in terms
of the free oscillator modes $\gamma^{\mu}_{m}$ (\r{19}). Then
a straight forward calculation, involving $XX$ OPE and contour
arguement yields the NC commutator given in (\r{26}), thereby reproducing the results of \cite{chu, chu1, rb, br1, jing}, even though we have made use
of newly proposed normal ordering \cite{br} which is compatible with BCs.


\end{document}